\documentclass{ws-ijmpa}
\newif\ifbw\bwtrue
\newcommand{\reaktion}{\mbox{$pp\to \,pp\:\!K^+\!K^- $ }}
\begin{document}

\markboth{W.~Oelert for the COSY-11 Collaboration}
{ }

%
\catchline{}{}{}{}{}
%

\title{ General thoughts to the Kaon pair 
production in the threshold region}

\author{\footnotesize W.~Oelert$^{\%}$$^,$\footnote{E-mail address: 
w.oelert@fz-juelich.de}~, 
H.-H.~Adam$^{\#}$, 
A.~Budzanowski$^{\$}$,
E.~Czerwi\'nski$^{\star}$,
R.~Czy\.zykiewicz$^{\star}$,
D.~Gil$^{\star}$,
D.~Grzonka$^{\%}$, 
M.~Janusz$^{\star}$, 
L.~Jarczyk$^{\star}$, 
B.~Kamys$^{\star}$, 
A.~Khoukaz$^{\#}$, 
P.~Klaja$^{\star}$,  
P.~Moskal$^{\star}$, 
C.~Piskor-Ignatowicz$^{\star}$, 
J.~Przerwa$^{\star}$, 
T.~Ro\.zek$^{+}$, 
R.~Santo$^{\#}$, 
T.~Sefzick$^{\%}$, 
M.~Siemaszko$^{+}$, 
J.~Smyrski$^{\star}$, 
A.~T\"aschner$^{\#}$, 
P.~Winter$^{\times}$,  
M.~Wolke$^{\%}$, 
P.~W\"ustner$^{\otimes}$, 
W.~Zipper$^{+}$, 
}

\address{
$^{\%}$IKP, Forschungszentrum J\"ulich, J\"ulich, Germany\\ 
$^{\#}$IKP, Westf\"alische Wilhelms-Universit\"at, M\"unster, Germany\\
$^{\$}$Institute of Nuclear Physics, Cracow, Poland\\ 
$^{\star}$Institute of Physics, Jagellonian University, Cracow, Poland\\ 
$^{+}$Institute of Physics, University of Silesia, Katowice, Poland\\
$^{\times}$Department of Physics, University of Illinois at Urbana-Champaign, Urbana, IL 61801 USA\\
$^{\otimes}$ZEL, Forschungszentrum J\"ulich, J\"ulich, Germany\\
}

\maketitle

\begin{abstract}
Simple--minded thoughts about the cross sections for the reactions
{\mbox{$pp\to \,pp\:\!K^+\!K^- $ }} and {\mbox{$pp\to \,pp\:\!K^0\!K^0 $ }}
are presented, which certainly do not account for the complex coupled channel
problem but rather provide some ideas into the mutual reaction dynamics.
\end{abstract}
\section{Introduction}  
\begin{figure}[htp]
\begin{center}
\parbox[c]{0.6\textwidth}{
\centering
\includegraphics[width=0.6\textwidth]{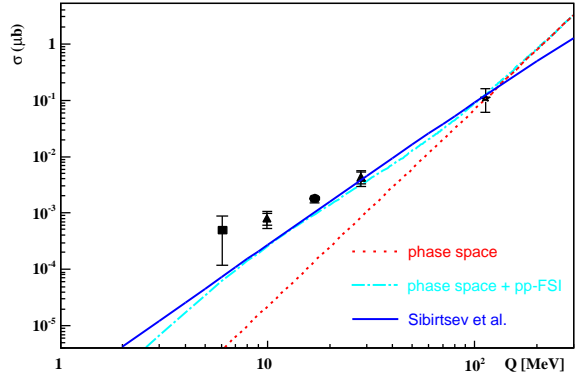}}
\vspace*{8pt}
\caption{Total cross section as a function of the excess energy $Q$ for the 
reaction {\mbox{$pp\to \,pp\:\!K^+\!K^- $ }}~
\protect \cite{Winter,QUENT,WOLKE,DISTO}.  
}
\label{excitation}
\end{center}
\end{figure}

Total cross sections~\cite{Winter,QUENT,WOLKE,DISTO} ~of the reaction \reaktion 
at excess energies below $Q=120$\,MeV are given in Figure~\ref{excitation}.
At low excess energies the data points lie significantly 
above the various expectations indicated by the different lines as long as 
these predicted curves
are all normalized to the DISTO point~\cite{DISTO} at $Q=114\,$MeV. 
The pure 
phase space (dashed line) differs by two orders of magnitude 
at $Q=10\,$MeV and a factor of five to ten at $Q=28\,$MeV. In comparison to that, the 
inclusion of the $pp$-FSI (\ifbw dashed-dotted \else red solid \fi line) by folding its 
parameterization known from the three body final state with the four body phase space 
is already closer to the experimental results but does not fully account for the difference. 
The solid line representing the calculation within a one-boson exchange 
model~\cite{sibirtsev:97} reveal a similar discrepancy as the $pp$-FSI parameterization. 
This model includes an energy dependent scattering amplitude derived from the fit of the 
total cross sections in $K^\pm p \to K^\pm p$ \cite{baldini:88} while the $pp$-FSI was 
not included, yet. Up to now, there is no full calculation available but the new data 
demands for further theoretical efforts in order to give a complete picture of 
the $K^+\!K^-$ production.
\section{The model} 
An important aspect might be the mass splitting between the neutral 
$K^0\!\bar{K^0}$
and 
charged $K^+\!K^-$
kaons being in the order of 8\,MeV. Based on the theoretical observation
that the opening neutral kaon production channel shows a substantial influence 
on the $\pi\pi \to K^+\!K^-$ transition (c.f. Figure 2. in 
reference \cite{krehl:97}), we tried a 
simple--minded Ansatz for the energy dependence of the excitation 
function for the \reaktion reaction assuming that with the 
opening of the neutral kaon channel (at 8\,MeV excess energy) some yield is taken
out of the charged kaon channel. As long as the total energy is large enough to
produce the charged kaon pair but is below the neutral kaon channel all 
strength for the associated strangeness production is devoted to the 
$K^+\!K^-$ creation. At 8\,MeV excitation energy the charged channel faces the
competition of the neutral one. 
\begin{figure}[htp]
\begin{center}
\parbox[c]{0.65\textwidth}{
\centering
\includegraphics[width=0.65\textwidth]{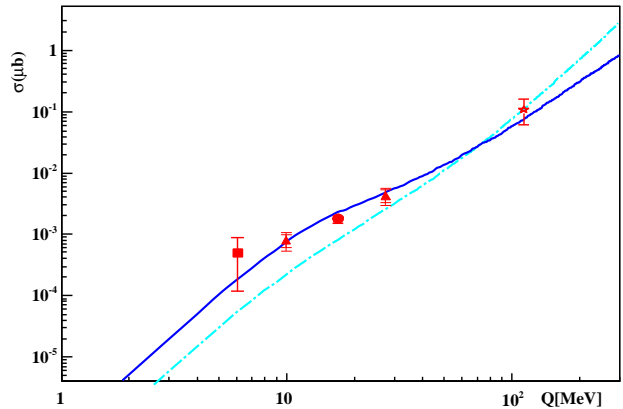}}
\vspace*{8pt}
\caption{Experimental data together with a simplified Ansatz for the 
incorporation the $K^0\!\bar{K}^0$ channel (solid line).}
\label{knullknull}
\end{center}
\end{figure}
\section{Cross section extraction}
For estimating such a coupled channel effect, 
we postulated the 
two simple assumptions. First, we assume that the excitation function for the
neutral $pp\to pp K^0\!\bar{K}^0$ channel follows exactly the \ifbw dashed-dotted \else red 
solid \fi line in figure \ref{excitation} for the charged kaon production case,
but shifted by 8\,MeV. 
If $f(Q)$ describes the excitation function of
the \ifbw dashed-dotted \else red solid \fi line, we simply assume that the 
energy dependence for the neutral channel is given by $g(Q) = f(Q-8\,$MeV)
where $Q$ refers here to the $K^+\!K^-$ system. 
Second, the opening channel is incorporated by the idea that the modified 
description $\bar{f}(Q)$ of the \reaktion channel is given by a subtraction 
of the neutral channel via $\bar{f}(Q):=c \cdot [f(Q)-g(Q)]$ with an 
arbitrary normalization $c$. This is certainly an extreme scenario since 
the influence of the $K^0\!\bar{K}^0$ channel is assumed to be 
an uncorrelated sum of the $K^+\!K^-$ and $K^0\!\bar{K}^0$ cross sections;
and, since the inverse transition $K^0\!\bar{K}^0 \to K^+\!K^-$
is not considered to take place.
The resulting energy 
dependence $\bar{f}(Q)$  as shown in Figure~\ref{knullknull}  just happens to 
pass through the experimental data. 
 
This good agreement should be taken with caution since a full coupled channel 
calculation has to be performed to determine quantitatively the effect of the 
$K^0\!\bar{K}^0$ channel, which  is expected to be rather small~\cite{HAID} 
and can not account for the full enhancement seen in the data.
Furthermore other effects are not considered which should
certainly be taken fully into account such
as final state interactions between the subsystems $p - K^+, ~p - K^-$ and 
$K^+ - K^-$ as well as the influence of intermediate resonances. \\
\\
We acknowledge the support of the
European Community-Research Infrastructure Activity
under the FP6 "Structuring the European Research Area" programme
(HadronPhysics, contract number RII3-CT-2004-506078),
of the FFE grants (41266606 and 41266654) from the Research Centre J{\"u}lich,
of the DAAD Exchange Programme (PPP-Polen),
of the Polish State Committe for Scientific Research
(grant No. PB1060/P03/2004/26), and of the
RII3/CT/2004/506078 - Hadron Physics-Activity -N4:EtaMesonNet.

\end{document}